\documentclass[journal=jacsat,manuscript=article]{achemso}

\usepackage[version=3]{mhchem} 

\author{Sotirios Fragkos}
\email{sotirios.fragkos@u-bordeaux.fr}
\affiliation{Universit\'e de Bordeaux - CNRS - CEA, CELIA, UMR5107, F33405 Talence, France}

\author{Evgenia Symeonidou}
\affiliation{Institute of Nanoscience and Nanotechnology, National Center for Scientific Research ‘Demokritos’, 15310 Athens, Greece}
\alsoaffiliation{School of Chemistry, Aristotle University of Thessaloniki, 54124 Thessaloniki, Greece}

\author{Emile Lasserre}
\affiliation{Universit\'e de Bordeaux - CNRS - CEA, CELIA, UMR5107, F33405 Talence, France}

\author{Baptiste Fabre}
\affiliation{Universit\'e de Bordeaux - CNRS - CEA, CELIA, UMR5107, F33405 Talence, France}

\author{Dominique Descamps}
\affiliation{Universit\'e de Bordeaux - CNRS - CEA, CELIA, UMR5107, F33405 Talence, France}

\author{Stéphane Petit}
\affiliation{Universit\'e de Bordeaux - CNRS - CEA, CELIA, UMR5107, F33405 Talence, France}

\author{Polychronis Tsipas}
\affiliation{Institute of Nanoscience and Nanotechnology, National Center for Scientific Research ‘Demokritos’, 15310 Athens, Greece}

\author{Yann Mairesse}
\affiliation{Universit\'e de Bordeaux - CNRS - CEA, CELIA, UMR5107, F33405 Talence, France}

\author{Athanasios Dimoulas}
\affiliation{Institute of Nanoscience and Nanotechnology, National Center for Scientific Research ‘Demokritos’, 15310 Athens, Greece}

\author{Samuel Beaulieu}
\email{samuel.beaulieu@u-bordeaux.fr}
\affiliation{Universit\'e de Bordeaux - CNRS - CEA, CELIA, UMR5107, F33405 Talence, France}

\title{Excited States Band Mapping and Ultrafast Nonequilibrium Dynamics in Topological Dirac Semimetal 1T-ZrTe$_2$}

\abbreviations{Ultrafast Dynamics, Topological Materials, Photoemission}
\keywords{American Chemical Society, \LaTeX}

\begin{document}

\begin{abstract}
We performed time- and polarization-resolved extreme ultraviolet momentum microscopy on topological Dirac semimetal candidate 1T-ZrTe$_2$. Excited states band mapping uncovers the previously inaccessible linear dispersion of the Dirac cone above the Fermi level. We study the orbital texture of bands using linear dichroism in photoelectron angular distributions. These observations provide hints on the topological character of 1T-ZrTe$_2$. Time-, energy- and momentum-resolved nonequilibrium carrier dynamics reveal that intra- and inter-band scattering processes play a capital role in the relaxation mechanism, leading to multivalley electron-hole accumulation near the Fermi level. We also show that electrons' inverse lifetime has a linear dependence as a function of their excess energy. Our time- and polarization-resolved XUV photoemission results shed light on the excited state electronic structure of 1T-ZrTe$_2$ and provide valuable insights into the relatively unexplored field of quantum-state-resolved ultrafast dynamics in 3D topological Dirac semimetals. 
\end{abstract}

\textbf{Keywords:} Ultrafast Dynamics, Photoemission, Topological Materials, Dichroism, Electronic Structure, Molecular Beam Epitaxy. 


\section{Introduction} 
Three-dimensional (3D) topological semimetals, including Dirac, Weyl, and nodal-line semimetals, have garnered significant interest, due to their unique electronic structures characterized by gapless topologically and symmetry-protected crossings between valence and conduction bands~\cite{Armitage18, Lv21}. In Dirac semimetals, these crossings are protected by the crystal rotational symmetry. They are formed through band inversion along the \textit{k$_z$} direction of the three-dimensional Brillouin zone~\cite{Yang14}. 

Angle-resolved photoemission spectroscopy (ARPES) is widely considered the most advanced technique for examining the electronic structure of solids. As a result, ARPES has become a tool of choice for studying topological materials, which feature characteristic non-trivial bulk and surface states, e.g. topological surface states and Fermi arcs, which can be revealed using this technique~\cite{Xia2009-by, Liu2014-sm, Xu2015-hr, Xu2015-dq, Lv2019-la, Krieger2024-he}. The use of ultrafast laser pulses enables a new flavor of ARPES measurements (time-resolved ARPES - trARPES) that allows for the investigation of excited states and ultrafast dynamics in materials~\cite{Boschini24}. Although this technique has been extensively used on various topological materials, including Weyl semimetals~\cite{Dai15, Crepaldri17, Hein20, Beaulieu2021}, topological insulators~\cite{Sobota12, Wang12, Reimann14, Zhu2015-eo, Herb2024-vg}, and graphene~\cite{Gierz2013-ps, Gierz15, Choi2024-sm, Merboldt2024-cn}, its application to 3D Dirac fermions has been limited~\cite{Bao2022-zz, Lin2024-sp}. 

1T-ZrTe$_2$ was recently proposed to be a topological Dirac semimetal with a type-II crossing along the \textit{k$_z$} direction~\cite{Fragkos2021-de}. Previous static ARPES studies had provided the first experimental evidence that few-layer 1T-ZrTe$_2$ epitaxially grown by molecular beam epitaxy (MBE) is a 3D topological Dirac semimetal~\cite{Tsipas2018-ii}. More specifically, it was shown that linearly dispersing bands along $\Gamma$-K and $\Gamma$-M directions cross near the Fermi level, indicating the existence of Dirac fermions, even down to the ultimate 2D limit of a single monolayer. In these measurements, the Dirac point was shown to be located at —or very close to— the Fermi level, which is notably different from the theoretical calculations that predict its location relatively far above it~\cite{Tsipas2018-ii, Fragkos2021-de, Kar2020-za, Nguyen22}. Despite these preliminary results on the electronic structure of 1T-ZrTe$_2$, its 3D topological Dirac semimetallic character has been subject to debate. Indeed, synchrotron ARPES studies on single crystals of Cr-doped 1T-ZrTe$_2$~\cite{Zhang2020-pp} did not reveal clear Dirac-like features, unlike epitaxially grown thin films~\cite{Tsipas2018-ii}. However, other studies combining synchrotron ARPES on 1T-ZrTe$_2$ single crystals with DFT calculations suggested that it is a Dirac semimetal~\cite{Kar2020-za}, but without direct observation of the Dirac crossing. Fermiology studies using de Haas–van Alphen oscillations have shown that 1T-ZrTe$_2$ exhibits no signature of non-trivial Berry phase~\cite{Nguyen22}. This observation was attributed to the fact that the type-II Dirac cone could lie far above the Fermi level ($\sim$500 meV)~\cite{Nguyen22}. On the other hand, recent magnetotransport measurements pointed towards the topological nature of 1T-ZrTe$_2$~\cite{Ng2020-pj, Wang2019-wg, Ou2022-pg}, and a recent theoretical study using topological quantum chemistry predicted that 1T-ZrTe$_2$ is a topological Dirac semimetal~\cite{Bradlyn2017-dg}. However, since the Dirac crossing is most probably lying above the Fermi level, accessing these subtle features with static ARPES poses significant challenges. Excited state band mapping using trARPES represents a potential solution to this issue.

\begin{figure}[t]
\begin{center}
\includegraphics[width=8.6cm]{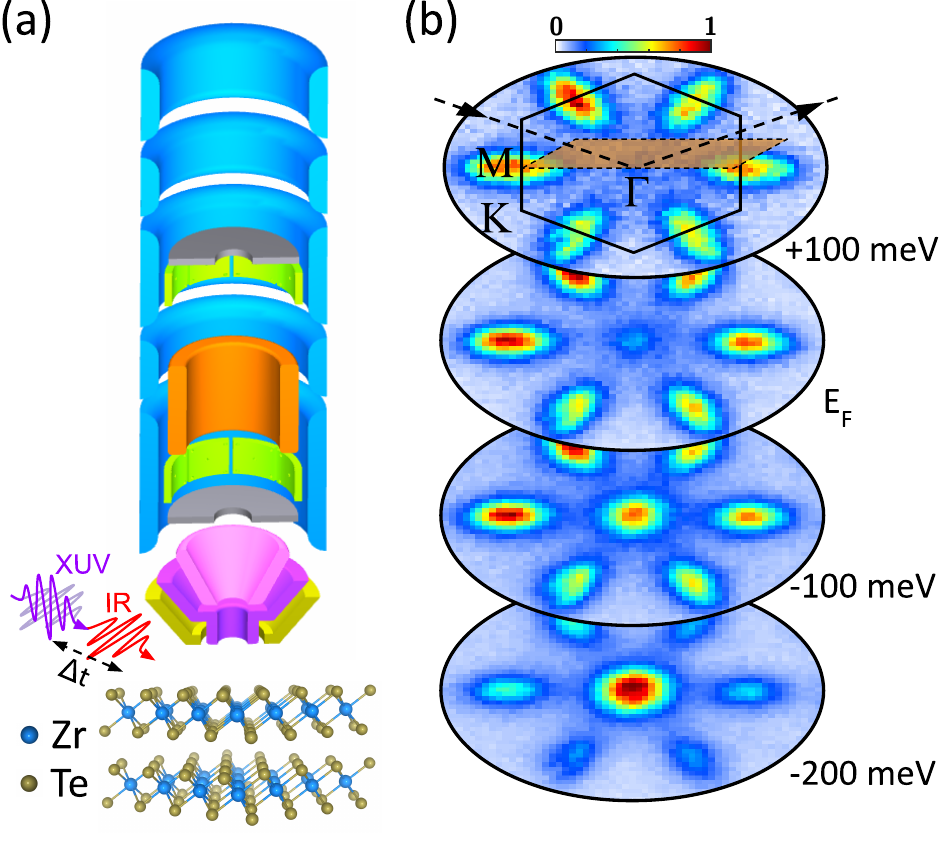}
\caption{\textbf{Schematic of the experiment.} \textbf{(a)} An infrared pump and a polarization-tunable (\textit{s}- or \textit{p}-polarized) XUV (21.6~eV) probe pulses are focused onto an MBE-grown 1T-ZrTe$_2$ sample, in the interaction chamber of a time-of-flight momentum microscope, at an incidence angle of 65$^{\circ}$. \textbf{(b)} Constant energy contours for different binding energies, obtained by summing photoemission intensities measured with \textit{s}- and \textit{p}-polarized XUV. These measurements were taken at the pump-probe temporal overlap, with an s-polarized IR pump pulse (1.2~eV, 135~fs, 2.95 mJ/cm$^2$). The light scattering plane (orange tilted rectangle overlaid with CEC in \textbf{(b)}) is aligned with the crystal mirror plane $\mathcal{M}$, i.e. is along the $\Gamma$-M direction.}
\label{fig:schematic}
\end{center}
\end{figure}

In this study, we perform time- and polarization-resolved extreme ultraviolet (XUV) momentum microscopy on the 1T-ZrTe$_2$ topological Dirac semimetal candidate. By photoexciting electrons into high-lying conduction bands using infrared pump pulses (1030~nm - 1.2~eV), we uncover the energy-momentum dispersion of the electronic band structure above the Fermi level. Furthermore, the linear dichroism in photoelectron angular distributions (LDAD) obtained from polarization-resolved measurements on the photoexcited 1T-ZrTe$_2$ allows us to reveal different orbital textures of bands in an extended energy-momentum range, typically inaccessible using conventional dichroic ARPES measurements. Second-derivative analysis of the photoemission intensity shows that the Dirac cone crosses the conduction band $\sim$250 meV above the Fermi level. Moreover, energy- and momentum-resolved ultrafast carrier dynamics reveal that photoexcited electrons' inverse lifetime has a linear dependence on the excess energy. Our time- and quantum-state-resolved ultrafast electrons and holes dynamics measurements allow us to unveil microscopic processes, including intra- and inter-valley scattering, governing the nonequilibrium dynamics in photoexcited 3D topological Dirac semimetals. 

\section{Results} 
Figure~\ref{fig:schematic}(a) shows a schematic of the experimental geometry. An \textit{s-}polarized infrared (IR, 1.2~eV, 135~fs, 2.95 mJ/cm$^2$) pump and polarization-tunable (\textit{s}- or \textit{p}-polarized) XUV (21.6~eV) probe pulses are superimposed onto the 1T-ZrTe$_2$ sample at room temperature in the interaction chamber of the time-of-flight momentum microscope. The 1T-ZrTe$_2$ samples were grown by molecular beam epitaxy (MBE) on Si(111)/InAs(111) substrates. More information about the experimental setup and sample preparation can be found in the Supporting Information and elsewhere~\cite{Comby22, tkach2024multimode, Tsipas2018-ii}. 

ZrTe$_2$ belongs to the 1T transition metal dichalcogenides layered octahedral family with space group \textit{P}$\bar{3}$\textit{m}1 (No. 164). It is anticipated that the 3D gapless Dirac crossing originates from two bands associated with different irreducible representations, which prohibits gap opening due to hybridization. Theoretical calculations have predicted a pair of fourfold degenerate Dirac nodes $\sim$500~meV above the Fermi level, at ($k_x$=0, $k_y$=0, $k_z$=±0.228\textit{c}*) momenta~\cite{Fragkos2021-de}, where \textit{c}* is the inverse out-of-plane lattice constant. These Dirac nodes are protected by \textit{C$_3$} rotational symmetry along the \textit{c} axis. Figure~\ref{fig:schematic}(b) also shows constant energy contours for different binding energies, obtained by summing photoemission intensities obtained with \textit{s}- and \textit{p}-polarized XUV, at pump-probe overlap with the s-polarized pump pulse.

\begin{figure}[H]
\begin{center}
\includegraphics[width=8.6cm]{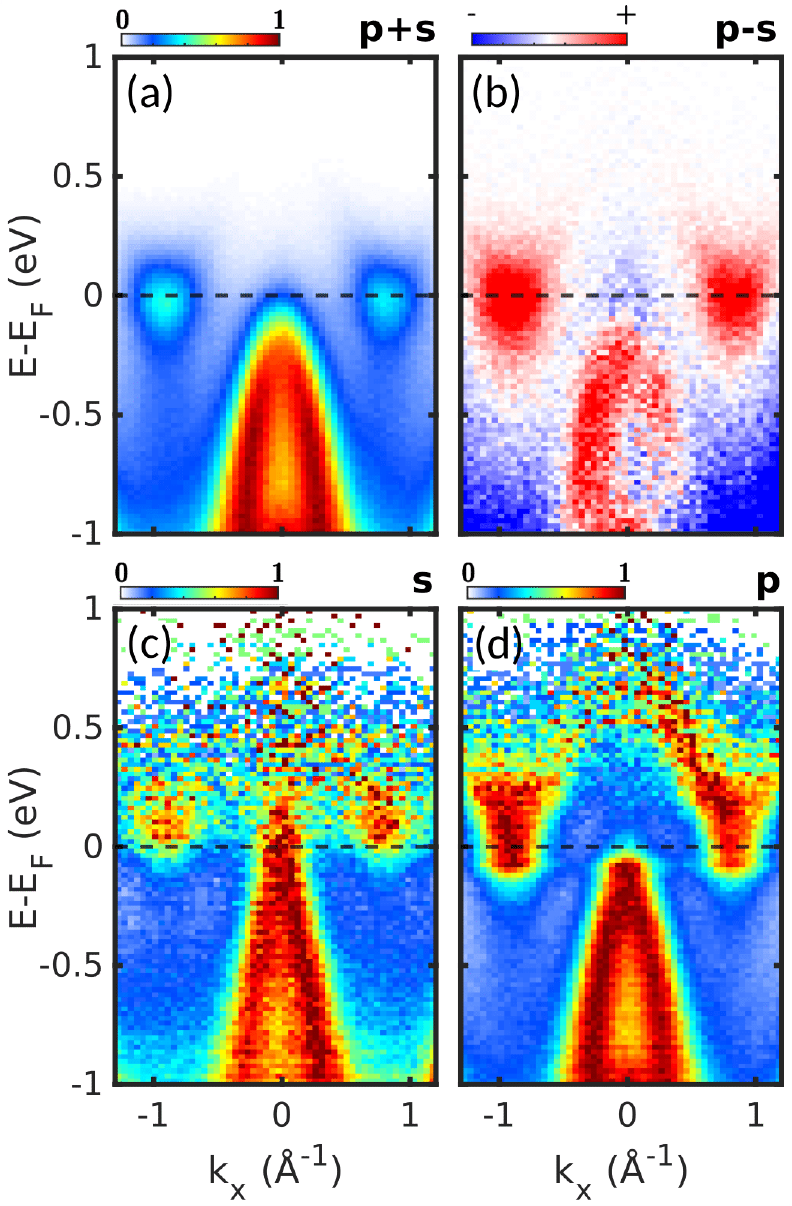}
\caption{\textbf{Linear Dichroism in Photoexcited 1T-ZrTe$_2$.} Band structure mapping of 1T-ZrTe$_2$ along M-$\Gamma$-M direction, at pump-probe overlap (\textit{$\Delta$t} = 0 ps), using s-polarized pump pulse to avoid spurious effect of laser-assisted photoemission. \textbf{(a)} Raw signal (unnormalized) obtained by summing photoemission intensities measured using s- and p-polarized XUV probe pulse. \textbf{(b)} XUV LDAD extracted by subtracting the photoemission intensities measured using p- and s-polarized XUV probe pulses. \textbf{(c)} Normalized photoemission intensity measured using s-polarized XUV probe pulses. \textbf{(d)} Normalized photoemission intensity measured using p-polarized XUV probe pulses. In \textbf{(c)}-\textbf{(d)}, each momentum-resolved energy slice of the photoemission intensity is normalized by its maximum value.}
\label{fig:ldad}
\end{center}
\end{figure}

The energy- and momentum-dependence of the photoemission transition dipole matrix element holds valuable information about wavefunctions underlying the electronic band structure. Notable examples include the orbital character of the electronic band structure~\cite{Wang12, Sterzi2018-bq, Bentmann21, Sam2020, Sam2021-op} as well as Berry curvature~\cite{Cho18, Schuler2020-lw, Cho2021-rp, Schuler2022-bm}, which are probed by the modification of photoemission intensities upon the modulation of polarization-state of the ionizing radiation, in a scheme referred to as dichoic ARPES. In that spirit, we performed polarization-resolved measurements at pump-probe overlap (\textit{$\Delta$t} = 0 fs), by measuring the energy- and momentum-resolved photoemission intensity while continuously rotating the XUV linear polarization axis angle. A cut of the photoemission intensity along M-$\Gamma$-M, integrated over all XUV polarization states, is shown in Fig.~\ref{fig:ldad}(a). Figures~\ref{fig:ldad}(c)-(d) shows the normalized (for each energy) photoemission intensities along the same high-symmetry cut, for \textit{s-} and \textit{p-}polarized XUV pulses. One can notice striking differences between these two configurations. Indeed, with \textit{s}-polarized XUV, a linear Dirac band extending a few 100 meV above the Fermi level is probed around $\Gamma$, while with \textit{p-}polarized XUV, the Dirac band is absent, and a parabolic hole-like band appears. Additionally, the contribution of the electron-like pockets at M points is strongly enhanced with \textit{p}-polarized XUV. Previous polarization-dependent synchrotron-ARPES study on 1T-ZrTe$_2$ has shown similar photoemission intensity modulation upon changing polarization and attributed this behavior to the different orbital character of the bands below Fermi level~\cite{Kar2020-za}. Additionally, the orbital-resolved band structure calculations reported in this study predicted that the parabolic band at $\Gamma$ and electron-like M pockets are mostly composed of orbitals with out-of-plane characters (Te \textit{p$_z$} and Zr \textit{d$_{z^2}$}, respectively), while the tip of the Dirac cone by Zr \textit{d$_{xz}$} and \textit{d$_{yz}$} orbitals. The XUV LDAD presented in Fig.~\ref{fig:ldad}(b) is in good agreement with the above-mentioned predictions. Indeed, at strong off-normal geometry (angle of incidence of 65$^{\circ}$), p-polarized light is expected to favor photoemission from out-of-plane orbitals. In addition, our pump-probe scheme also allows accessing LDAD of states above the Fermi level, which is not possible using static polarization-resolved ARPES. This enables us to extract simultaneously the LDAD from the M pockets,  the parabolic band at $\Gamma$ (red color), and the tip of the linear Dirac cone dispersing above the Fermi level around $\Gamma$ point (blue color). It is thus evident that the photoemission transition dipole matrix elements significantly influence the ARPES spectra of 1T-ZrTe$_2$, and that \textit{s}-polarized XUV substantially enhances the photoemission intensity from the Dirac cone.

\begin{figure}[t]
\begin{center}
\includegraphics[width=8.6cm]{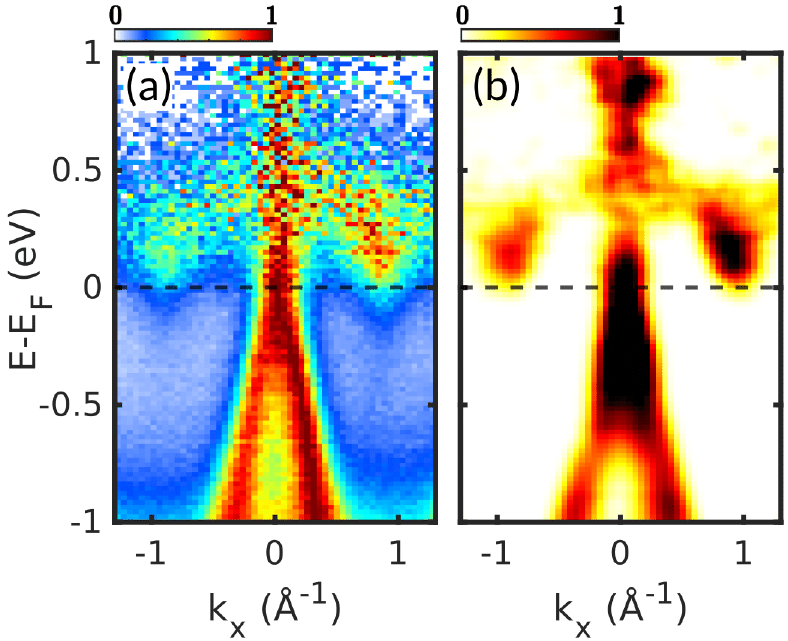}
\caption{\textbf{Excited State Band Mapping.} \textbf{(a)} Normalized energy-momentum cut along M-$\Gamma$-M direction at pump-probe overlap (\textit{$\Delta$t} = 0 ps). The IR pump and the XUV probe are \textit{s}-polarized. \textbf{(b)} Second derivative analysis of the energy-momentum cut along M-$\Gamma$-M direction shown in \textbf{(a)}. The second derivative presented in \textbf{(b)} is calculated by summing the second derivatives along the momentum and energy axes. More details about the second derivative procedure can be found in the Supporting Information.}
\label{fig:mapping}
\end{center}
\end{figure}

To gain a deeper insight into the dispersion of the Dirac cone, high-statistic measurement was conducted at pump-probe overlap ($\Delta t = 0$ fs) with \textit{s}-polarized XUV (Fig.~\ref{fig:mapping}(a)). The associated photoemission intensity along the M-$\Gamma$-M direction shows a clear Dirac-like cone extending far above the Fermi level. Second derivative analysis of this signal (Fig.~\ref{fig:mapping}(b)) (More details about the second derivative procedure are given in the Supporting Information, Fig. S2) indicates that the electron-like conduction bands with minima at M points disperse upward towards the $\Gamma$ point, merging with the tip of the Dirac cone at $\sim$250 meV above the Fermi level. An electron-like band at $\Gamma$ can also be seen above this crossing (around 0.6~eV$<$$\mathrm{E-E_F}$$<$1~eV). These observations are in good agreement with previously reported calculations, which predict that 1T-ZrTe$_2$ is a 3D topological Dirac semimetal~\cite{Tsipas2018-ii, Fragkos2021-de, Kar2020-za, Nguyen22}. It is important to note that the Dirac bands also exhibit dispersion along the \textit{k$_z$} axis, and accurate determination of the position of the Dirac point in the 3D Brillouin zone would require photon-energy-dependent~\cite{Liu2014-sm, Xu2015-hr, Yan2017-kx} trARPES measurements. While this is currently not possible with our setup, time-resolved ARPES with XUV photon-energy tunability are now emerging~\cite{Sie2019, Guo22,  Heber22, Chen23, Majchzak24, majchrzak2024}, and could be very useful to probe the 3D dispersion of electronic states above the Fermi level. However, this type of experiment represents a major challenge. 

\begin{figure}[t]
\begin{center}
\includegraphics[width=8.6cm]{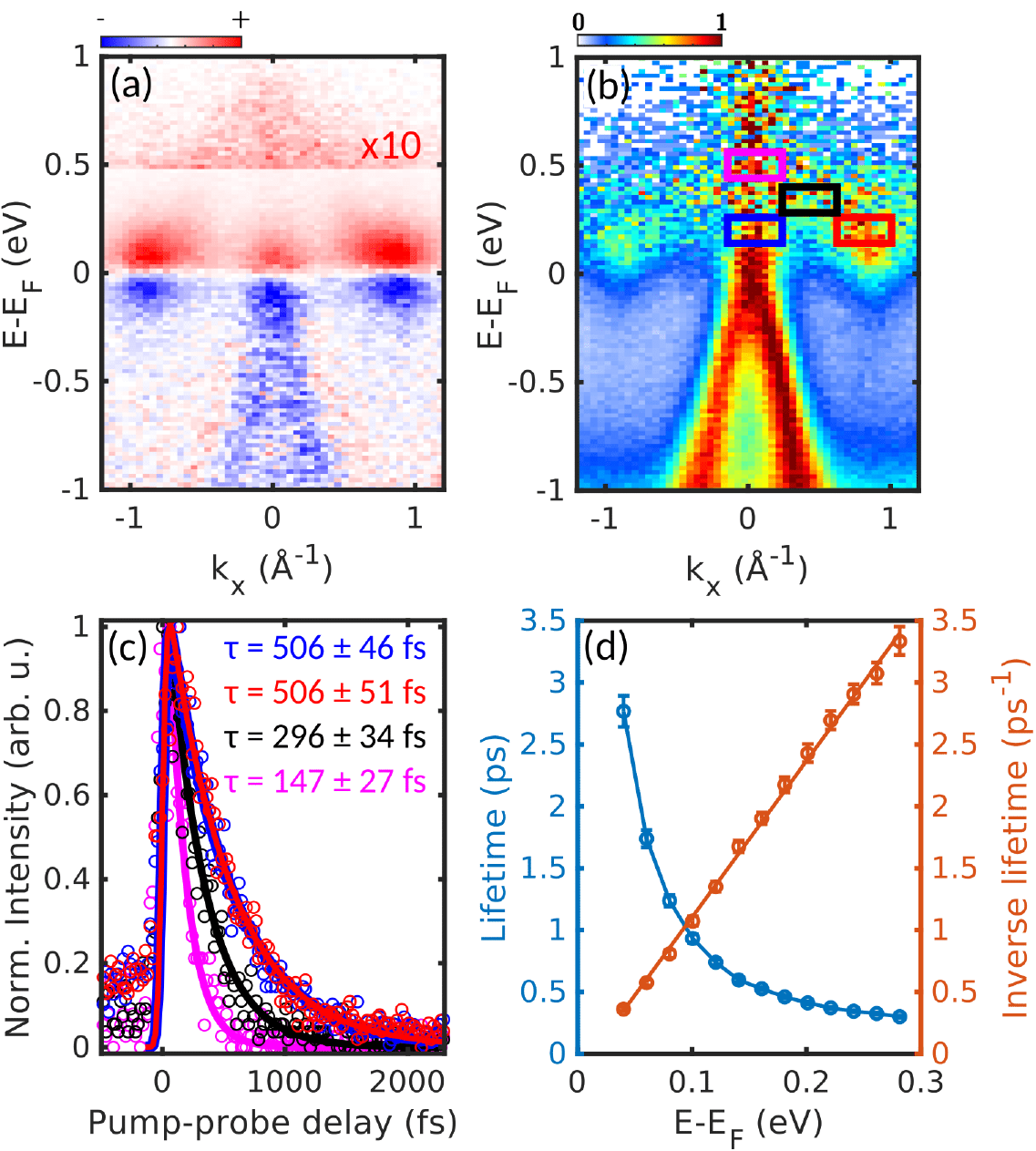}
\caption{\textbf{Energy- and Momentum-Resolved Ultrafast Nonequilibrium Dynamics.} \textbf{(a)} Energy- and momentum-resolved differential map, generated by subtracting the photoemission intensity obtained before (integration between -1.25 ps and -0.95 ps) and during/after (integration between 0 ps and +0.3 ps) optical photoexcitation. This differential map was obtained by summing photoemission intensities for all three  M-$\Gamma$-M directions of the Brillouin zone to enhance statistics. \textbf{(b)} Energy-momentum cut along M-$\Gamma$-M direction with integration boxes used in \textbf{(c)}, where energy- and band-resolved temporal evolution of the photoexcited electrons within these regions of interest are shown. \textbf{(d)} Momentum-integrated electron population lifetimes and associated inverse lifetimes as a function of excess energy.}
\label{fig:ultrafast}
\end{center}
\end{figure}

After meticulously analyzing the excited states dispersion and orbital characters, we now focus on the ultrafast nonequilibrium carrier dynamics in photoexcited 1T-ZrTe$_2$. We performed pump-probe measurements with \textit{s}-polarized IR pump, to avoid spurious effects of laser-assisted photoemission during the optical cross-correlation, and \textit{s}-polarized XUV pulses to efficiently access the ultrafast dynamics of carriers within the Dirac cone. Figure~\ref{fig:ultrafast}(a) shows the energy- and momentum-resolved differential map along M-$\Gamma$-M high-symmetry directions, generated by subtracting the photoemission intensity obtained before (integration between -1.25 ps and -0.95 ps) and during/after (integration between 0 ps and +0.3 ps) optical photoexcitation. Negative differential photoemission intensity (blue color) mainly results from the hole creation upon optical photoexcitation and subsequent intra- and inter-band scattering processes. Conversely, positive differential photoemission intensity (red color) mainly originates from electron creation upon optical photoexcitation and subsequent scattering processes. Excitation-induced band renormalizations can also leave their imprint onto these differential maps~\cite{Crepaldri17, Hein20, Beaulieu2021}. However, under these experimental conditions, we did not observe any significant band renormalization between the equilibrium and photoexcited states of the sample.

In the differential map (Fig.~\ref{fig:ultrafast}(a)), we observe a very faint positive signal centered around $\Gamma$ and extending up to $~\sim$1~eV above the Fermi level. This indicates that vertical optical transitions induced by the IR pump occur between the Dirac cone and the high-lying excited states around $\Gamma$. These hot electrons can subsequently relax through intra- and interband scattering and accumulate near the Fermi level within the tip of the Dirac cone at $\Gamma$ and within M pockets, leading to a strong positive (red) signal. In addition, we observe significant hole accumulation just below the Fermi level, both within the Dirac cone at $\Gamma$ and within M pockets. The creation of holes at M pockets most likely arises from phonon-assisted M-$\Gamma$ intervalley scattering, since there is no pair of bands allowing direct optical transition with a 1.2~eV photon near M points.

The temporal evolution of photoexcited electrons within various energy-momentum regions of interest (ROI - squares in Fig~\ref{fig:ultrafast}(b)) is depicted in Fig~\ref{fig:ultrafast}(c), revealing the energy- and band-resolved ultrafast electron dynamics. Surprisingly, the extracted lifetime  ($\sim$500~fs) at the tip of the Dirac cone above the Fermi level at $\Gamma$ (blue ROI box) is the same as the one extracted at the M pocket (red ROI box), at the same energy (between 0.14 eV and 0.26 eV), which implies that the carrier lifetime seems to be momentum-independent. This suggests that interband scattering processes that are faster than our temporal resolution ($\sim$140~fs) are at play, leading to an apparent momentum-independent carrier lifetime. For higher energy electrons (between 0.44 eV and 0.56 eV, pink ROI box), the extracted lifetime decreases to $\sim$150 fs. In addition, looking at time-resolved photoemission intensity in regions of interest symmetric in energy with respect to the Fermi level (below and above) reveals remarkably similar decay dynamics for electrons and holes (Fig. S3). This charge compensation between electron and hole pockets, even in nonequilibrium settings, was also observed in type-II Weyl semimetallic WTe$_2$ and was argued to be a key effect in explaining the nonsaturating magnetoresistance of this type of material~\cite{Caputo18}. 

Having established the momentum-independent behavior of ultrafast carrier lifetime, we can now extract the momentum-integrated (integrated for all momenta), energy-resolved lifetime of photoexcited electrons (Fig.~\ref{fig:ultrafast}(d)). For electrons near the Fermi level (E-E$_F$ = 40 meV), the population lifetime is relatively long, i.e. $\sim$3 ps, while for high-energy electrons (E-E$_F$ = 280 meV), the lifetime is as small as $\sim$300 fs. Indeed, the inverse lifetime increases linearly as a function of excess energy, with a slope of (12.6 $\pm$ 0.4 ps$^{-1}$eV$^{-1}$), as one can see in Fig.~\ref{fig:ultrafast}(d). This linear dependence of the inverse lifetime as a function of excess energy was already observed in other semimetals~\cite{Bao2022-zz, Lin2024-sp}, where the electron-phonon coupling strength \textit{g} was derived from this slope (\textit{S}), using the equation $S={g(T_e-T_L)/(2C_e k_BT_e)}$, where \textit{C$_e$} is the electron heat capacity, \textit{T$_e$} is the transient electronic temperature, and \textit{T$_L$} is the lattice temperature. Determining electron-phonon coupling strength \textit{g} using energy-resolved lifetime measurements thus requires knowledge about the electron heat capacity \textit{C$_e$}, which, to the best of our knowledge, has yet not been reported for 1T-ZrTe$_2$. Moreover, this unconventional non-Fermi-liquid behavior of the electron inverse lifetime as a function of excess energy has been predicted in layered materials, such as graphite~\cite{Gonz96}. This phenomenon arises from the combined effects of linearly dispersing bands near the Fermi level and reduced Coulomb screening. A similar behavior has also been observed in the excitonic insulator candidate Ta$_2$NiSe$_5$, which features a quasi-linearly dispersing conduction band~\cite{mor22}. In this material, the non-Fermi-liquid electron inverse lifetime as a function of excess energy was also shown to increase with excitation density due to transient screening~\cite{mor22}. In the case of 1T-ZrTe$_2$, low dimensionality and reduced screening likely play a significant role in determining the photoexcited electron lifetime.

Even though we cannot extract electron-phonon coupling strength from our energy-resolved lifetime measurements, we can compare the ultrafast carrier dynamics in 1T-ZrTe$_2$ with similar measurements performed on other topological semimetals. In the Dirac nodal line semimetal PtSn$_4$, an ultrafast population lifetime as short as 400 fs was observed for electrons 100~meV above the Fermi level. This short population lifetime results from multichannel electron-phonon scattering among its complex metallic bands~\cite{Lin2024-sp}. In PtSn$_4$, the slope extracted from the inverse lifetime as a function of excess energy is around $\sim$14 ps$^{-1}$eV$^{-1}$~\cite{Lin2024-sp}, a value similar to the one we obtained for 1T-ZrTe$_2$. Conversely, in 3D type-I Dirac semimetal Cd$_3$As$_2$, a significantly longer carrier lifetime of $\sim$3 ps for electrons 100 meV above the Fermi level, and a slope of $\sim$2 ps$^{-1}$eV$^{-1}$ (inverse lifetime as a function of excess energy) was measured~\cite{Bao2022-zz}. It was argued that carrier lifetime in Cd$_3$As$_2$ is significantly longer than in e.g. 2D Dirac fermions in graphene, because of lower optical phonon energy. In 1T-ZrTe$_2$, at the same energy (100 meV above Fermi level), the electron lifetime is $\sim$1 ps, i.e. in between lifetimes measured for PtSn$_4$ and Cd$_3$As$_2$. To explain this behavior, scattering phase space arguments can be used. Indeed, type-I Dirac semimetals have nearly zero density of states (DOS) at the Dirac point, which strongly restricts scattering phase space for electron-hole recombination. This can explain the long lifetime observed in Cd$_3$As$_2$ and is also at the origin of population inversion in this material, i.e. the Dirac point acting as a bottleneck for the ultrafast carriers relaxation~\cite{Bao2022-zz}. In 1T-ZrTe$_2$, the ultrafast decay mechanism is more complex, since neighboring (non-Dirac) bands also allow for interband scattering processes, similar to the case of PtSn$_4$. Moreover, a trARPES study on MoTe$_2$ and WTe$_2$ type-II Weyl semimetals reported an ultrafast population lifetime of $\sim$465 fs and $\sim$1180 fs, respectively, for electrons slightly above the Fermi level~\cite{Crepaldri17}. These fast population lifetimes, on the same order of magnitude as the ones obtained for PtSn$_4$ and 1T-ZrTe$_2$, likely originate from the relatively large electron and hole DOS near the Fermi level. Therefore, the shorter lifetime, and the absence of population inversion, can be associated with larger phase space available for electron-hole scattering and recombination. In addition, these observations could also be linked to the nature of the topological crossing in 1T-ZrTe$_2$. As mentioned above, 1T-ZrTe$_2$ is predicted to be a type-II Dirac semimetal~\cite{Fragkos2021-de, Kar2020-za, Nguyen22}. In contrast to type-I, type-II Dirac and Weyl semimetals have a non-zero DOS at the Dirac/Weyl points, since they appear as the touching point between strongly tilted electron and hole pockets. Therefore, a type-II Dirac crossing could not act as a bottleneck for the relaxation of photoexcited electrons.

\begin{figure}[t]
\begin{center}
\includegraphics[width=8.6cm]{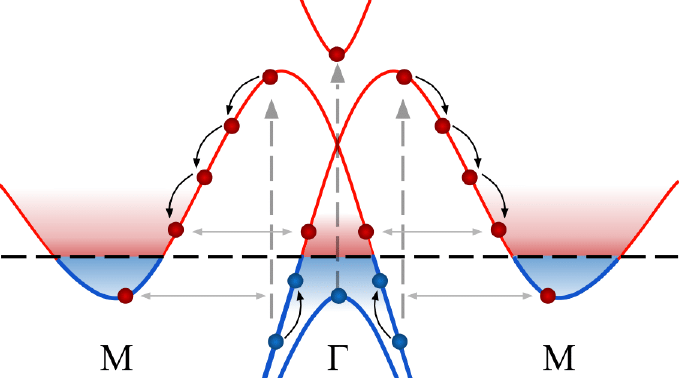}
\caption{\textbf{Schematic of Ultrafast Nonequilibrium Carrier Dynamics in 1T-ZrTe$_2$}. The vertical dashed grey lines represent optical transitions leading to initial electrons and holes creation. The horizontal light grey arrow represents interband transitions, while the curved black arrows represent intravalley scattering. Shaded red and blue areas represent the accumulation of electrons and holes, above and below the Fermi level.}
\label{fig:nonlin}
\end{center}
\end{figure}

In Fig.~\ref{fig:nonlin}, we summarize the main steps underlying the ultrafast carrier dynamics in 1T-ZrTe$_2$, which were revealed using our time- and polarization-resolved momentum microscopy measurements. The IR pump pulse induces vertical interband transitions (vertical dashed grey arrows), creating photoholes (small blue circles) in the band manifold near $\Gamma$ below the Fermi level, and excited electrons (small red circles) in high-lying bands at associated momentum. This photoexcitation triggers a cascade of intra- and inter-band scattering events that bring the system back towards equilibrium. Indeed, hot electrons can scatter to lower energies, filling the tip of the Dirac cone at $\Gamma$ and M pockets near the Fermi level. Concomitantly, intervalley scattering between $\Gamma$ and M bands allows reshuffling the hole population below the Fermi level. Multivalley electron-hole recombination can thus bring back the electronic system to equilibrium, on a few picosecond timescale. 

\section{Conclusions} 
In conclusion, our time- and polarization-resolved XUV momentum microscopy of 1T-ZrTe$_2$ allowed us to reveal the previously inaccessible linear dispersions of the bulk Dirac cone above the Fermi level. The linear dichroism in photoelectron angular distributions (LDAD) from photoexcited 1T-ZrTe$_2$ enabled us to unveil different orbital textures of bands below and above the Fermi level, which is beyond the reach of conventional dichroic ARPES measurements. Excited states band mapping combined with linear dichroism hints at the topological character of the electronic states of this material. We determined the position of the Dirac point, which is located $\sim$250 meV above the Fermi level. Ultrafast time-, energy- and momentum-resolved photoemission measurements using well-chosen IR pump and XUV probe polarization states allowed us to provide a quantum-state-resolved picture of scattering processes governing the out-of-equilibrium behavior of this topological Dirac semimetal. We showed that both intra- and inter-band scattering processes play a capital role in ultrafast carrier relaxations, leading to multivalley electron-hole accumulation near the Fermi level. In addition, the momentum-integrated inverse lifetime shows linear dependence, as in other Dirac and nodal-line semimetals. Our work sheds light on the previously experimentally unmapped excited states of 1T-ZrTe$_2$ and provides insights into the relatively unexplored field of ultrafast dynamics in 3D topological Dirac semimetals. 

\begin{acknowledgement}

We thank Nikita Fedorov, Romain Delos, Pierre Hericourt, Rodrigue Bouillaud, Laurent Merzeau, and Frank Blais for technical assistance. We thank Gerd Schönhense, Olena Tkach, and Sergey Babenkov for stimulating discussion regarding momentum microscopy. We acknowledge the financial support of the IdEx University of Bordeaux / Grand Research Program "GPR LIGHT". We acknowledge the funding from MSCA-ITN project SMART-X-860553. We acknowledge support from ERC Starting Grant ERC-2022-STG No.101076639, Quantum Matter Bordeaux, AAP CNRS Tremplin and AAP SMR from Université de Bordeaux. Funded by the European Union. Views and opinions expressed are however those of the author(s) only and do not necessarily reflect those of the European Union. Neither the European Union nor the granting authority can be held responsible for them.

\end{acknowledgement}

\begin{suppinfo}

In the Supporting Information, additional discussions and/or figures about the experimental setup, growth methodology, decapping process, band structure mapping and 2$^{nd}$ derivative analysis, balanced ultrafast electron and hole dynamics, as well as momentum-independent carriers' lifetime analysis are given.

\end{suppinfo}

\section{Data Availability Statement}

The data that support the findings of this article are openly available on Zenodo 
 \url{https://doi.org/10.5281/zenodo.17224698}

\section{Supporting Information}

\subsection{Experimental Setup}
The time-resolved momentum microscope apparatus at Centre Lasers Intenses et Applications (CELIA) is centered around a custom-built, polarization-tunable ultrafast Extreme UltraViolet (XUV) beamline~\cite{Comby22}. We utilize a commercial high-repetition-rate (166~kHz) ytterbium (Yb) fiber laser operating at 1030 nm with a pulse duration of 135 fs (FWHM) and an output average power of 50 W (Amplitude Laser Group). For the XUV probe arm, a portion of the beam is frequency-doubled using a beta barium borate (BBO) crystal to generate 515 nm pulses. The resulting beam is converted into an annular beam by a drilled mirror (5 W average power - 30 $\mu$J/pulse) and tightly focused into a thin, dense argon gas jet to drive high-order harmonic generation. The intense annular 515 nm driver is spatially filtered out with pinholes while its 9th harmonic (21.6 eV central photon energy) is spectrally selected using a combination of three Sc/SiC multilayer XUV mirrors under quasi-normal incidence (NTTAT) and transmission through a 200 nm thick Sn metallic filter. The polarization axis direction of the driver, and consequently of the XUV beam, is adjustable by rotating a half-wave plate positioned before the high-order harmonic generation chamber. The first XUV mirror is spherical and collimates the XUV beam. The last XUV mirror focuses the beam onto the sample, down to a 35$\mu$m $\times$ 45$\mu$m spot size.  For the pump arm, a minor fraction (typically 60 mW) of the fundamental infrared (IR) laser beam is employed. The IR pump beam (1030 nm, 1.2eV central photon energy) is focused by a f=80~mm lens, and its 70$\mu$m $\times$ 140$\mu$m focal spot is superimposed to the XUV probe beam by a drilled mirror under vacuum. Photoemission data are collected using a custom time-of-flight momentum microscope (GST mbH)\cite{tkach2024multimode}, enabling simultaneous detection of electrons across the full surface Brillouin zone over an extensive binding energy range, without requiring sample reorientation\cite{Medjanik17}. The samples are mounted on a motorized six-axis hexapod for precise alignment within the interaction chamber of the momentum microscope (base pressure of 2×10$^{-10}$ mbar). The emitted photoelectrons are imaged by a set of electrodes and collected by microchannel plates and a delay-line detector, which enables determining their position and time-of-flight. Post-processing of the data is performed using an open-source workflow~\cite{Xian20, Xian19_2}, which efficiently converts the raw single-event-based datasets into binned, calibrated data hypervolumes of the desired dimensions.

\subsection{Growth Methodology}
\begin{figure}[t]
\begin{center}
\includegraphics[width=8.6cm]{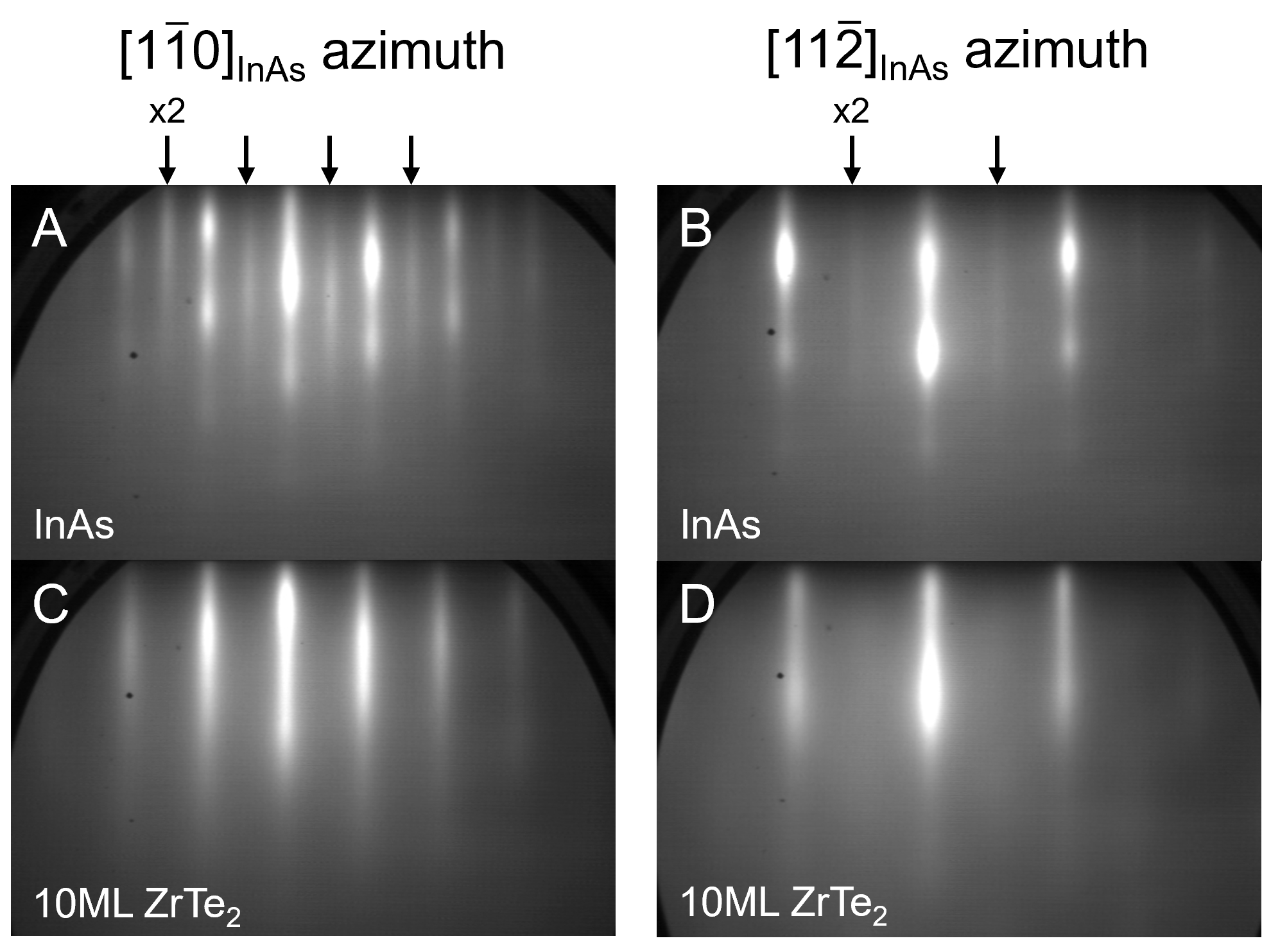}
\caption{\textbf{Reflection High Energy Electron Diffraction} RHEED patterns of \textbf{(a)}-\textbf{(b)} InAs(111) substrate with the 2$\times$2 reconstructed surface streaks, and of \textbf{(c)}-\textbf{(d)} 1T-ZrTe$_2$ multilayer film.  \textbf{(a)} and \textbf{(c)} are RHEED partterns along the InAs[1$\bar{1}$0], while  \textbf{(b)} and \textbf{(d)} are along [11$\bar{2}$] azimuths. The streaky pattern of ZrTe$_2$ film \textbf{(c)}-\textbf{(d)} indicates a smooth, well-ordered surface without rotational domains.}
\label{fig:rheed}
\end{center}
\end{figure}

1T-ZrTe$_2$ multilayers (20 nm thickness) are epitaxially grown on Si(111)/InAs(111) substrates. Before growth, InAs(111)/Si(111) substrates are chemically cleaned \textit{ex-situ} in 5-6N HCl in isopropanol for 5 minutes, to etch the surface oxide. The substrates are subsequently rinsed with isopropanol for 30 seconds to avoid reoxidation. An \textit{in-situ} annealing step at 430$^{\circ}$C in UHV follows, to get a clean and flat, In-terminated InAs(111) surface, evidenced by a 2$\times$2 reconstruction ~\cite{Taguchi2006-bi} in RHEED (Fig.~\ref{fig:rheed}). Epitaxy takes place in an ultra-high vacuum molecular beam epitaxy (MBE) vertical chamber (with a base pressure of $\approx$ 5$\times$10$^{-9}$ Torr), under Te-rich conditions. Elemental Zr (99.95\%, slug) and Te (99.999$\%$, granule) are used as precursors. The growth temperature is set at $\approx$ 450$^{\circ}$C, while the growth rate is fixed at 0.05 monolayer/s. A 50nm-thick Te capping is deposited on top of 1T-ZrTe$_2$ samples to prevent oxidation. 

The substrate and epilayer surface ordering and reconstruction are monitored using \textit{in-situ} RHEED. After the cleaning process, a 2×2 reconstruction of the InAs(111) surface is observed (figure~\ref{fig:rheed} (a)-(b)), which is expected for a clean, oxygen-free, In-terminated InAs(111) surface ~\cite{Taguchi2006-bi}. The streaky pattern of ZrTe$_2$ film (figure~\ref{fig:rheed} (c)-(d)) indicates a smooth, well-ordered surface without rotational domains. Despite the significant lattice mismatch (6.6$\%$) between the epilayer and the substrate ~\cite{Tsipas2018-ii}, the ZrTe$_2$ film is rotationally aligned in-plane with respect to the InAs(111) substrate, a characteristic of vdW epitaxial growth.

\subsection{Decapping Proccess}
A 50 nm Te capping deposition occurred \textit{in-situ} by MBE to prevent oxidation of the 1T-ZrTe$_2$ multilayers. The Te decapping process of 1T-ZrTe$_2$ multilayers takes place at a base pressure of 1$\times$10$^{-9}$ mbar. We sputter with Ar$^{+}$ at V = 600V, an emission current of 10 mA (using an extractor type ion source optimized for sample cleaning -- IQE 11/35, SPECS GmbH), and Ar overpressure of 8 $\times$ 10$^{-6}$ mbar, introduced using a leak valve, for 3 min. Following the sputtering, we annealed the sample for 60 min at 380$^{\circ}$C. After capping removal, the samples were introduced in a motorized 6-axis hexapod for sample alignment in the main trARPES chamber (base pressure of 2$\times$10$^{-10}$ mbar).

\subsection{Band structure mapping and 2$^{nd}$ derivative analysis}
\begin{figure}[H]
\begin{center}
\includegraphics[width=8.4cm]{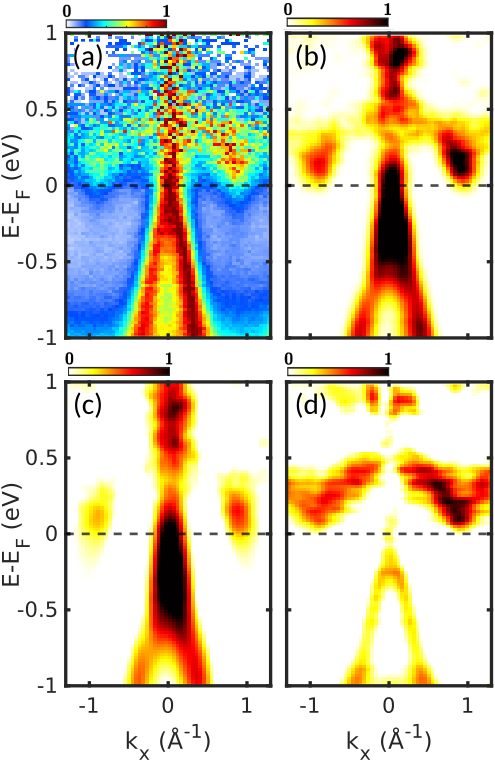}
\caption{\textbf{Band structure mapping and 2$^{nd}$ derivative analysis} \textbf{(a)} Normalized energy-momentum cut along M-$\Gamma$-M direction at pump-probe overlap (\textit{$\Delta$t} = 0 ps). Both the IR pump and the XUV probe are\textit{s}-polarized. \textbf{(b)} Second derivative analysis of the energy-momentum cut along M-$\Gamma$-M direction shown in \textbf{(a)}. The second derivative presented in \textbf{(b)} is calculated by summing the second derivatives along the momentum \textbf{(c)} and energy axes \textbf{(d)}.}
\label{fig:second}
\end{center}
\end{figure}

To gain a deeper insight into the dispersion of the Dirac cone, we performed a second derivative analysis of the energy-momentum cut along M-$\Gamma$-M direction measured at pump-probe overlap (\textit{$\Delta$t} = 0 ps) using \textit{s}-polarized IR pump and \textit{s}-polarized XUV probe (Fig.~\ref{fig:second}(a)). As mentioned in the main text, second derivative analysis indicates that the electron-like conduction band with minima at M merges with the tip of the Dirac cone $\sim$250 meV above the Fermi level at $\Gamma$ point. The second derivative shown in Fig.~\ref{fig:second}(b) is obtained by summing up the second derivative of the spectra along the momentum (Fig.~\ref{fig:second}(c)) and the energy axes (Fig.~\ref{fig:second}(d)). Second derivative analysis along the momentum axis, shown in Fig.~\ref{fig:second}(c), captures well the strongly dispersive Dirac cone at $\Gamma$ point. However, this method does not capture the less-dispersive bands at the M points. To address this, we also perform a second derivative analysis along the energy axis, which better captures the dispersion of the M pockets. Finally, by summing the two-second derivative spectra, we reveal that the Dirac cone intersects the electron-like conduction bands at the $\Gamma$ point (Fig.~\ref{fig:second}(b)).

\hspace{5mm}

\subsection{Balanced ultrafast electron and hole dynamics}

\begin{figure}[H]
\begin{center}
\includegraphics[width=8.6cm]{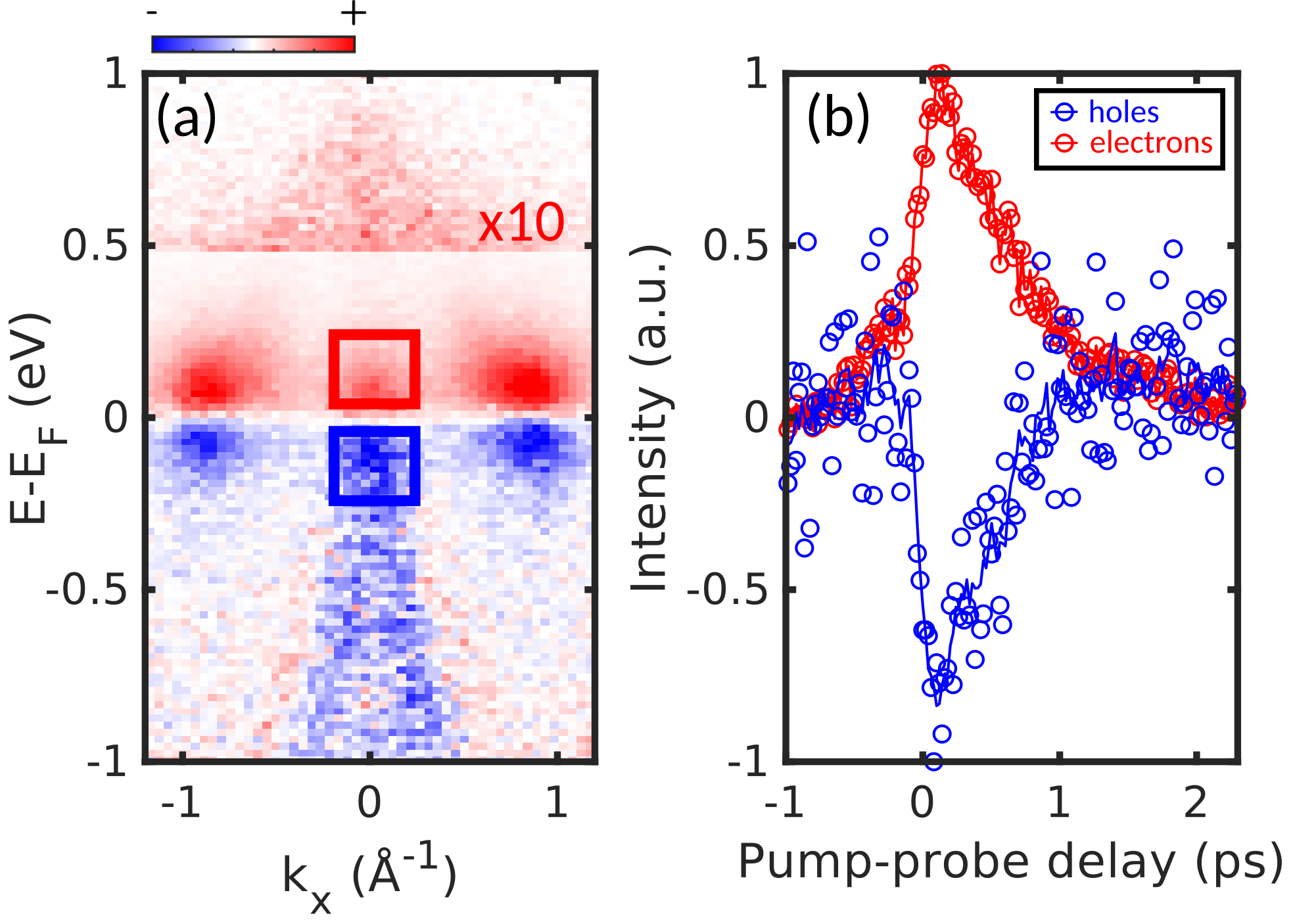}
\caption{\textbf{Balanced ultrafast electron and hole dynamics} \textbf{(a)} Energy- and momentum-resolved differential map, generated by subtracting the photoemission intensity obtained before (integration between -1.25 ps and -0.95 ps) and during/after (integration between 0 ps and +0.3 ps) optical photoexcitation. This differential map was obtained by summing photoemission intensities for all three  M-$\Gamma$-M directions of the Brillouin zone to enhance statistics. The red and blue boxes correspond to the integration windows used in \textbf{(b)} to obtain the temporal evolution of electrons and holes.}
\label{fig:balanced}
\end{center}
\end{figure}

In this section, we investigate the time-resolved photoemission intensity in regions of interest symmetric in energy with respect to the Fermi level (below and above). The temporal evolution of electrons and holes within designated regions of interest (shown by squares in Fig.~\ref{fig:balanced}(a) and positioned symmetrically above and below the Fermi level at $\Gamma$) is shown in Fig.~\ref{fig:balanced}(b). This analysis reveals remarkably similar decay dynamics for electrons and holes.  

\subsection{Momentum-independent lifetime}

\begin{figure}[H]
    \centering
    \includegraphics[width=0.7\textwidth]{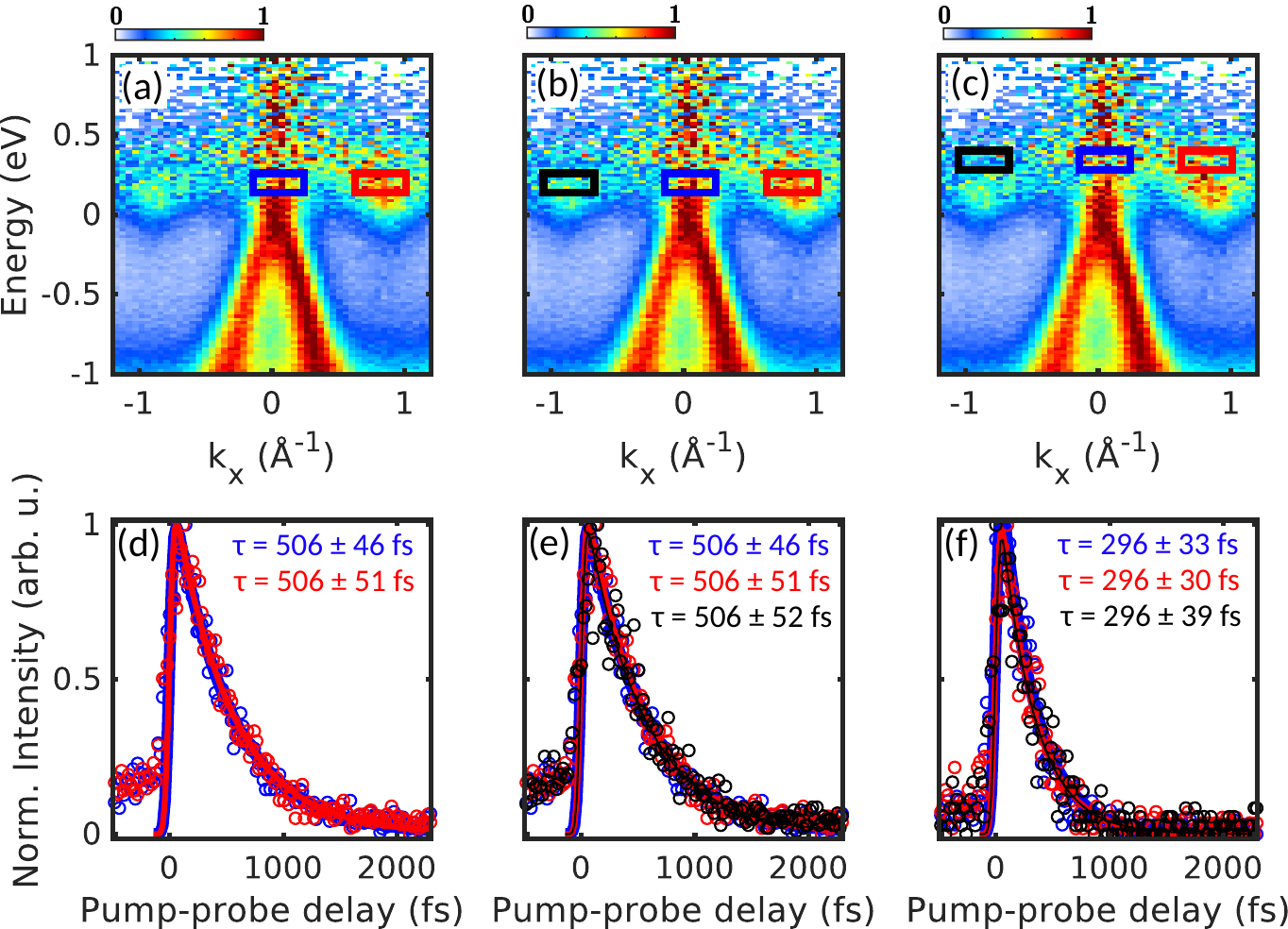}
    \caption{\textbf{Energy- and Momentum-Resolved Ultrafast Nonequilibrium Dynamics.} \textbf{(a-c)} Energy-momentum cut along M-$\Gamma$-M direction with integration boxes used in \textbf{(d-f)}, where energy- and band-resolved temporal evolution of the photoexcited electrons within these regions of interest are shown.}
    \label{fig:dyn}
\end{figure}

In this subsection, we investigate the temporal evolution of photoexcited electrons within various energy-momentum regions of interest (ROIs), indicated by boxes in Fig.\ref{fig:dyn}(a-c). As discussed in the main text, the extracted lifetime at the tip of the Dirac cone above the Fermi level at $\Gamma$ (blue ROI box) is approximately 500fs (Fig.\ref{fig:dyn}(d)). This lifetime is identical to that extracted at the M pocket (red ROI box) within the same energy range (between 0.14~eV and 0.26~eV), suggesting that the carrier lifetime is independent of momentum. In Fig.\ref{fig:dyn}(b), an additional region of interest (black ROI box) is highlighted at the M' pocket, at the same energy. As expected, the extracted lifetime (Fig.\ref{fig:dyn}(e)) is also $\sim$500fs. To further confirm the momentum-independent nature of the excited state lifetime, we shifted the ROIs by 120meV, as shown in Fig.\ref{fig:dyn}(c). This shift resulted in lifetimes of 296~fs for all three ROIs (Fig.~\ref{fig:dyn}(f)). This analysis reveals remarkably consistent decay dynamics for photoexcited electrons of the same energy, regardless of momentum.

\providecommand{\latin}[1]{#1}
\makeatletter
\providecommand{\doi}
  {\begingroup\let\do\@makeother\dospecials
  \catcode`\{=1 \catcode`\}=2 \doi@aux}
\providecommand{\doi@aux}[1]{\endgroup\texttt{#1}}
\makeatother
\providecommand*\mcitethebibliography{\thebibliography}
\csname @ifundefined\endcsname{endmcitethebibliography}  {\let\endmcitethebibliography\endthebibliography}{}

\end{document}